\documentclass[a4paper]{article}
\usepackage{INTERSPEECH2020}
\usepackage{graphicx, balance, color}
\usepackage{tabularx, booktabs, fancyvrb, pifont}
\usepackage{pifont, hyperref}

\newcolumntype{Y}{>{\centering\arraybackslash}X}
\newcolumntype{L}{>{\arraybackslash}X}
\newcolumntype{M}[1]{>{\raggedright\arraybackslash}m{#1in}}

\newcommand{\before}[1]{}

\title{North America Bixby Speaker Diarization System for the VoxCeleb Speaker Recognition Challenge 2021}
\name{Myungjong Kim, Taeyeon Ki, Aviral Anshu, Vijendra Raj Apsingekar}
\address{Samsung Research America, USA}
\email{\{myungjong.k, taeyeon.ki, aviral.anshu, v.akar\}@samsung.com}

\begin{document}

\maketitle

\begin{abstract}

This paper describes the submission to the speaker diarization track of VoxCeleb
Speaker Recognition Challenge 2021 done by North America Bixby Lab of
Samsung Research America.  Our speaker diarization system consists of four main
components such as overlap speech detection and speech separation,
robust speaker embedding extraction, spectral clustering with fused affinity matrix, and leakage filtering-based postprocessing.
We evaluated our system on the VoxConverse dataset and the
challenge evaluation set, which contain natural conversations of multiple
talkers collected from YouTube. Our system obtained 4.46\%, 6.39\%, and 6.16\%
of the diarization error rate on the VoxConverse development, test, and the
challenge evaluation set, respectively. 

\end{abstract}


\section{Introduction}

Speaker diarization is the process of labeling different speakers given an
audio stream, determining ``who spoke when" in a multi-speaker conversation~\cite{anguera2012speaker}.  
Speaker diarization has potential to be widely
utilized in a variety of applications such as meeting conversation analysis and
multi-media information retrieval.  It can be used as a front-end component of
automatic speech recognition (ASR), providing improved ASR accuracy, and more
rich analysis depending on speakers. 

In general, one of the most important components of speaker diarization is
speaker embedding extraction. Speaker embeddings are generally extracted from
short speech segments (e.g., 1.5 seconds) and directly used for speaker
clustering.  Therefore, extracting reliable speaker embeddings is critical in
achieving better speaker diarization. 

There are several challenges to extract better speaker embeddings in a
multi-speaker conversation in the real world.  Conversation often can happen in
adverse noisy environment and multiple speakers can speak at the same time.  In
this situation, speaker embeddings could have low capacity to represent speaker
identity.  To overcome this problem, integrating additional components such as
speech enhancement~\cite{sun2018speaker} and speech
separation~\cite{xiao2021microsoft} into speaker diarization has shown good
improvements in speaker diarization of natural  conversation.

In this paper, we propose an integrated speaker diarization system, consisting
of overlap speech detection and separation, robust speaker embedding
extraction, effective spectral clustering and fusion methods, 
and postprocessing to handle incorrectly clustered results.  We evaluate our system
on the VoxConverse dataset~\cite{chung2020spot}, which is used in VoxCeleb
Speaker Recognition Challenge 2020.
VoxConverse~\footnotemark~\footnotetext{https://github.com/joonson/voxconverse}
is a large-scale diarization dataset, collected from YouTube videos including
talk-shows, panel discussions, political debates and celebrity interviews.  The
organizer shared ground truth data of the dev \& test set of VoxConverse as a
valdiation set for this year.  The VoxConverse dev \& test set consists of 216 recordings (20.3 hours in total) and
232 recordings (43.5 hours in total), respectively.  The number of speakers in one recording varies
from 1 to 21.  We report diarization error rates (DERs) and Jaccard error rates
(JERs) using the scoring tool provided by the organizer~\footnotemark~\footnotetext{https://github.com/JaesungHuh/VoxSRC2021}
on the VoxConverse dev and test
set, respectively.
Our best system on the VoxConverse set is submitted to the challenge. 

\label{sec:introduction}

\section{System Description}
\label{sec:system_description}

\begin{figure*}
\centering
\includegraphics[width=\textwidth]{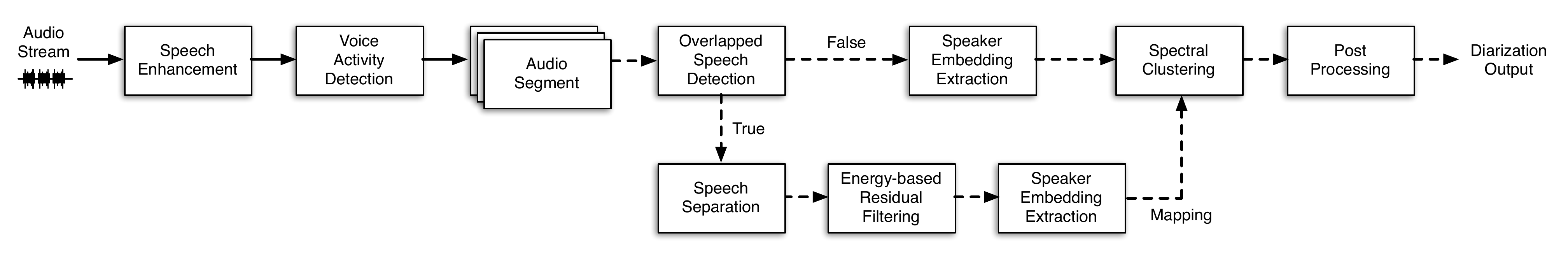}
\caption{System Diagram}
\label{fig:arch}
\end{figure*}

\autoref{fig:arch} shows a block diagram of our speaker diarization system. Each component is described in the following subsections.

\subsection{Speech enhancement}
Our system starts with preprocessing, which is a speech enhancement
technique to reduce background noise and improve speech quality, especially
retaining speaker specific information as much as possible. According to recent
research~\cite{sun2018speaker, landini2021analysis}, the long short-term memory
(LSTM) network-based speech enhancement method trained on simulated data
improves performance in speaker diarization. 
In our experiments, our diarization accuracy was also consistently improved by
3\% on average on the VoxConverse set, so we take advantage of this approach. 

\subsection{Voice activity detection}

We use the pretrained Silero voice activity detection (VAD) model~\cite{SileroVAD} as our baseline. The
model architecture is based on convolutional neural networks and transformers.
To improve voice detection accuracy in cases where our preprocessing does not remove
 diverse background or foreground noise properly, the pretrained model is retrained
with the AVA speech dataset~\cite{chaudhuri2018avaspeech}. The AVA speech dataset is
collected from YouTube, and contains 300 audios with 16000 clean and noisy speech,
and 17000 non-speech frames in a variety of environment. 

We have compared our retrained VAD model with the AVA speech data to the baseline model.
\autoref{tab:vad} shows missed speech (MS) rates and false alarm (FA) rates on
the VoxConverse dev and test sets.  As shown in \autoref{tab:vad}, the retrained VAD
model performs better than the baseline on both MS and FA. The error rates show
relative improvements of 5.45\% and 5.75\%, respectively.

\begin{table}
\caption{Comparison between the retrained VAD model and the pretrained Silero
  VAD model on the VoxConverse set.}
\label{tab:vad}
\centering
\begin{tabularx}{\columnwidth}{|Y|c|c|c|c|}
\hline
\textbf{Model} &
\multicolumn{2}{c|}{\textbf{Vox dev}} &
\multicolumn{2}{c|}{\textbf{Vox test}} \\
\cline{2-5}
&
\textbf{MS} &
\textbf{FA} &
\textbf{MS} &
\textbf{FA} 
\\
\hline
Baseline VAD & 2.96 & 1.02 & 3.02 & 2.08 \\
\hline
Retrained VAD  & 2.87 & 0.96 & 2.78 & 1.96 \\
\hline
\end{tabularx}
\end{table}


\subsection{Overlapped speech detection and separation}

In the natural daily conversations or meetings,
it is observed that people often speak together at the same time (i.e., overlapping utterances)~\cite{chung2020spot, ccetin2006analysis}.
However, a speaker embedding vector represents a single speaker's acoustic characteristics in general. 
So, handling a speech segment that contains multiple
speakers' speech is one of the major challenges in speaker diarization.  This
is because multiple speakers are mingled in a speech segment, and it is hard to
extract each speaker embedding vector correctly. In some cases, non-dominant
speaker's characteristics can be missing as well.


Neural network-based speech separation models such as ConvTasNet~\cite{abs-1809-07454},
Sepformer~\cite{subakan2021attention}, DPTNet~\cite{chen2020dualpath} and,
DPRNN~\cite{luo2020dualpath} have been widely studied, and show good
performance. However, the latency of speech separation still remains as an
issue. Thus, processing an entire audio signal with speech separation model is not
ideal.

To address this issue, we exploit pretrained overlapped speech detection (OSD) model~\cite{bullock2019overlapaware} 
to determine whether a segment has
multiple speakers' utterance. If it does, we conduct speech separation to find
each speaker's voice. Going into details, the OSD takes segments as input from
VAD, and splits into subsegments, 1.5 seconds long window with a shift of 0.75
seconds. The pretrained OSD model is trained with a subset of the AMI
corpus~\cite{carletta_unleashingthe}, recording in a quiet environment. We
have fine-tuned the hyper parameters of the model to make a good fit for
the VoxConverse dataset.

We conduct an experiment to see how well our fine-tuned OSD module works with
the VoxConverse dataset. We have extracted 100 overlapped and non-overlapped speech
segments from the VoxConverse dev set, and obtaind 75\% classification accuracy on
this set with our fine-tuning OSD model.

We consider two speech separation architectures to find a better fit for our
speaker diarization system in terms of latency and accuracy. One is
ConvTasNet~\cite{abs-1809-07454}, a fully-convolutional time-domain audio
separation network, and the other is Sepformer~\cite{subakan2021attention}, an
RNN-free Transformer-based neural network. Both models consist of three main
components including an encoder, a masknet, and a decoder.  Our ConvTasNet model
is trained with the combination of WSJ0-2, WHAMR!~\cite{Maciejewski2020WHAMR},
and VoxConverse datasets, but our Sepformer model is trained with only WHAM!~\cite{wichern2019wham}
dataset. From our experiment, regarding latency, ConvTasNet is a better option,
but Sepformer shows better performance on the VoxConverse dev and test sets.

We add a signal energy-based segment filter to our Sepformer model in order to
handle a false detection case, which is that the OSD classifies a single voice
in a segment into multiple voices. 
In the cases, we observed one of the separated signals often contains
the active speaker's voice while the other contains residual signals with low energy. 
Thus, this energy-based residual segment filtering helps to compensate a false detection case.
Also, we add one more VAD round based on
WebRTC with the most aggressiveness mode to cover the corner case, which is a short
partially overlapped speech in a subsegment. With the additional VAD round, the
non-speech part is filtered out in the separated signals.  

After speech separation, if a segment is determined as a single speaker 
(i.e., when the decision of energy-based residual filtering is true), we use the
original audio. This is because the speech acoustic characteristics can be
subject to change.



\begin{table*}
\caption{The DER (\%) and JER (\%) of the speaker diarization system.}
\label{tab:result}
\centering
\begin{tabularx}{\textwidth}{|L|c|c|c|c|c|c|c|}
\hline
\textbf{System} &
\multicolumn{2}{c|}{\textbf{Vox dev}} &
\multicolumn{2}{c|}{\textbf{Vox test}} &
\multicolumn{2}{c|}{\textbf{Challenge set}} \\
\cline{2-7}
&
\textbf{DER} &
\textbf{JER} &
\textbf{DER} &
\textbf{JER} &
\textbf{DER} &
\textbf{JER} 
\\
\hline
Baseline & - & - & - & - & 17.99 & 38.72 \\
\hline
ECAPA-TDNN & 5.02 & 22.02 & 6.76 & 32.04 & - & - \\
\hline
Light ECAPA-TDNN & 5.17 & 22.70 & 6.84 & 31.91 & -  & - \\
\hline
Light ECAPA-TDNN retrained & 5.18 & 22.72 & 6.82 & 32.31 & - & - \\
\hline
Fusion & 4.98 & 22.00 & 6.70 & 31.31 & - & - \\
\hline
Fusion + boosted affinity & 4.91 & 21.99 & 6.70 & 32.02 & - & - \\
\hline
Fusion + boosted affinity + leakage filtering & 4.82 & 21.90 & 6.46 & 31.90 & - & - \\
\hline
Fusion + boosted affinity + leakage filtering + overlap handling & 4.46 & 21.69 & 6.39 & 32.19 & 6.16 & 27.95 \\
\hline
\end{tabularx}
\end{table*}

\subsection{Speaker embedding extraction}

ECAPA-TDNN-based speaker embedding models were recently
introduced~\cite{desplanques2020ecapa}, and have shown a great success on speaker
verification~\cite{thienpondt2021idlab, thienpondt2020idlab} as well as
speaker diarization tasks~\cite{dawalatabad2021ecapa}. It is based on
TDNN-based x-vector speaker embeddings~\cite{snyder2018x} with several advancements by
incorporating a channel- and context-dependent attention mechanism in the
pooling layer, 1-dimensional Squeeze-Exitation (SE) blocks, 1-dimensional
Res2Net blocks, and multi-layer feature aggregation. In addition, the model is
optimized with the AAM-softmax loss~\cite{deng2019arcface} to effectively
classify speaker identities. Also, it directly optimizes the cosine distance
between the speaker embeddings, and therefore, it is beneficial to use cosine
similarities as a similarity measure during spectral clustering. 

We consider three ECAPA-TDNN models with slightly different variants: 1) large
version of ECAPA-TDNN, 2) light version of ECAPA-TDNN, and 3) light version of
ECAPA-TDNN with retraining. The large version of ECAPA-TDNN has the same structure as
in~\cite{desplanques2020ecapa, dawalatabad2021ecapa} with 80 dimensional log Mel
filterbank energies as acoustic features and TDNN channel size of 1024. This
model is trained on VoxCeleb 1~\cite{nagrani2017voxceleb} \&
2~\cite{chung2018voxceleb2} data with
RIRs~\footnotemark~\footnotetext{https://www.openslr.org/28/} and
MUSAN~\cite{snyder2015musan} datasets for data augmentation. We use the data
augmentation strategies~\cite{dawalatabad2021ecapa}: Waveform and Frequency
dropout, speech perturbation, reverberation, additive noise, and noise +
reverberation.  All corrupted data is added to our train set, and our training
data is six times the original data. To train the model, the data is cropped
into 3 seconds audios.  We used the Adam optimizer with a cyclical learning rate
using a triangular policy~\cite{smith2015cyclical}.  AAM-softmax is used with a
margin of 0.2.  Training is done for 10 epochs with batches of 32 segments.

In the light version of ECAPA-TDNN, acoustic features are replaced from 80
dimensional log Mel filterbank energies to 40, and TDNN channel size from 1024 to
512. This model is also trained on VoxCeleb data with the same data augmentation
strategies, but MUSAN is replaced with WHAMR! noise set. 
Inspired by~\cite{thienpondt2021idlab}, we retrain the light version of
ECAPA-TDNN model with 1.5 seconds audio clips to match the embedding window
size. In addition, we increased a margin of AAM-softmax from 0.2 to 0.5. 
Retraining is done for 2 more epochs based on the trained light version of ECAPA-TDNN.

All models contain an embedding layer with hidden unit size of 192 right before
the softmax layer. After speaker embedding model training, we extract 192
speaker embedding vectors from 1.5 seconds window with a shift of 0.75 seconds. 

\subsection{Spectral clustering}

Spectral clustering is a popular clustering method in speaker
diarization~\cite{wang2018speaker, park2021review}.  We apply the unnormalized
spectral clustering approach as in~\cite{dawalatabad2021ecapa}. The affinity
matrix is calculated using the cosine similarity. 
We prune out the smaller values in the affinity matrix to focus more on prominent values.
The symmetrized affinity matrix is used to estimate a unnormalized Laplacian matrix followed by eigendecomposition. 
The number of speakers k is estimated using the maximum eigengap approach and we use the first k eigenvectors. 
Specifically, the rows of the eigenvector matrix are the k dimensional spectral embeddings 
corresponding to each speaker embedding vector of a segment. 
Finally, we use the standard k-means algorithm to cluster the estimated spectral embeddings.

Affinity matrix directly contributes in obtaining spectral embeddings to be
clustered, therefore, manipulating the affinity matrix is one of the keys to achieve
better diarization results.  One widely used option is to prune out the smaller
values in affinity matrix as our default setting
~\cite{von2007tutorial}.  Another option is to binarize the similarity values by
converting non-zero values to 1 to have only 0 or 1 in the affinity matrix
~\cite{park2019auto} or normalize the similarity values after pruning out the
smaller values~\cite{wang2018speaker}. In our experiments, there was no
difference from these three methods in terms of DER. Instead, we add fixed values to the actual
cosine similarity to boost the values after pruning out the smaller values in affinity matrix.

Spectral clustering is only applied to non-overlapped segments. After obtaining
speaker clusters from non-overlapped speaker embeddings, the speaker embeddings
are mapped from overlapped segments (speech separation outputs). We use cosine
similarity between speaker embeddings and centroids of all speaker clusters, and 
the speaker embeddings are mapped to the cluster with the highest similarity.

\subsection{Postprocessing}

We perform postprocessing based on initial clustering results to obtain more
fine-grained results.  If the input audio includes non-stationary
background/foreground noise recorded in different acoustic setup or speech
separation signals, one speaker is likely to split into multiple clusters. To
overcome this issue, we compare cosine similarity between centroids of all
clusters, and if the similarity is larger than a threshold, we merge the
clusters. 

As noted in~\cite{xiao2021microsoft}, speech separation outputs often 
have residual noise signals in one of separated signals 
espceially when a single speaker is speaking with adverse background noises such as music.
In addition, VAD may incorrectly detect noise or music signals
as speech. To reduce the errors coming from the residual noises or misdetected
signals from VAD, we leverage leakage filtering~\cite{xiao2021microsoft} which
means if maximum cosine similarity of a segment to the centroids of all clusters
is below a threshold, the segment is removed from the diarization results.  In
addition, we further compute a centroid from the leakage filtered segments and
remove segments if similarity with a leakage centroid is larger than a
threshold. 
The threshold is set to 0.65, 0.2, and 0.7 which are tuned from the VoxConverse set, respectively.

\subsection{Affinity matrix fusion}

Three speaker embeddings are trained on slightly different configurations, and
therefore, different speaker characteristics may be captured on the short speech
segments to complement each other.  
Accordingly, affinity matrices might be estimated to focus on 
different views of relationship between speaker embeddings.
Spectral embeddings to be clustered is directly modeled from the affinity matrix, so it is important to obtain a reliable affinity matrix in achieving a better accuracy in speaker diarization. 
To this end, we combine affinity matrices estimated from the three speaker embedidngs by averaging them.

\section{Results}
\label{sec:Results}

We evaluated the proposed systems on the VoxConverse dev and test sets
as well as the challenge evaluation set in terms of diarization error rates
(DERs) and Jaccard error rates (JERs)~\cite{ryant2018first} in \autoref{tab:result}. 
The baseline is
the official baseline of the challenge shared by the organizer.  We first
compared the diarization results of single ECAPA-TDNN-based speaker embeddings.
The large version of ECAPA-TDNN was better than the light version. Two light
versions of ECAPA-TDNN were comparable, but the retrained ECAPA-TDNN showed
slight improvement in the DER on the VoxConverse test set.

By combining three affinity matrices generated from the above three speaker
embeddings, we obtained better accuracy than single speaker embeddings in terms
of both DER and JER. The fusion system and the boosted affinity matrix together
helped better improve performance. Leakage filtering helped contribute to
diarization results by reducing both the DER and JER as well. 

When we applied a series of components with fused and boosted affinity-based clustering, leakage
filtering, and overlap handling, we were able to obtain the best diarization
results, showing a DER of 4.46\% on the VoxConverse dev set and a DER of 6.39\%
on the VoxConverse test set.  Finally, we submitted this system to the
challenge and obtained a DER of 6.16\% and a JER of 27.95\%, achieving relative
improvements of 65.7\% and 27.8\% in the DER and JER, respectively, compared to
the challenge baseline system.

\section{Conclusions}

This paper describes the speaker diarization system submitted by North America
Bixby Lab of Samsung Research America.  The system consists of a series of
important components such as overlap speech detection and separation,
ECAPA-TDNN-based speaker embeddings, fused and boosted affinity matrix-based spectral
clustering, and leakage filtering-based postprocessing.  The proposed system was evaluated with
the data set provided by the VoxCeleb Speaker Recognition  Challenge 2021, and
we achieved DERs of 4.46\%, 6.39\%, and 6.16\% on the VoxConverse dev \& test
set and the challenge evaluation set, respectively. 

\label{sec:conclusions}

\balance 

\bibliographystyle{IEEEtran}
\bibliography{references}

\end{document}